\documentclass[a4paper, 11pt]{article}
\pdfoutput=1
\usepackage{jheppub}
\makeatletter
\def\@fpheader{}
\makeatother

\usepackage{graphicx,wrapfig,float,slashed}
\usepackage{amsmath,amssymb,dsfont,epsfig,graphicx}
\usepackage{mathtools} 
\usepackage{multirow}
\usepackage{blkarray}
\usepackage{epstopdf}
\usepackage{ragged2e}
\usepackage[utf8]{inputenc} 
\usepackage{cancel}
\usepackage[usenames,dvipsnames,table]{xcolor}
\usepackage[toc]{appendix}
\usepackage[normalem]{ulem}
\usepackage{enumitem}
\usepackage[font=small,skip=1pt]{caption}
\allowdisplaybreaks
\usepackage{pifont}
\usepackage{subcaption}
\usepackage{footnote}

\newcommand{\beq}{\begin{eqnarray}}
\newcommand{\eeq}{\end{eqnarray}}

\newcommand{\bea}{\begin{eqnarray}}
\newcommand{\eea}{\end{eqnarray}}
\newcommand{\del}{\partial}
\newcommand{\dd}{\ensuremath{\text{d}}}

\begin{document}
\title{Slowly rotating Q-balls}
\author[1,2]{Yahya Almumin,}
\emailAdd{yalmumin@uci.edu}
\author[3]{Julian Heeck,}
\emailAdd{heeck@virginia.edu}
\author[1]{Arvind Rajaraman,}
\emailAdd{arajaram@uci.edu}
\author[4]{and Christopher B. Verhaaren}
\emailAdd{verhaaren@physics.byu.edu}
\affiliation[1]{Department of Physics and Astronomy, \\
University of California, Irvine, CA 92697-4575, USA}
\affiliation[2]{Physics Department, Kuwait University,
 P.O. Box 5969 Safat, 13060, Kuwait}
 \affiliation[3]{Department of Physics, University of Virginia,
Charlottesville, Virginia 22904-4714, USA}
 \affiliation[4]{Department of Physics and Astronomy, Brigham Young University, Provo, UT, 84602, USA}

\abstract{
Q-balls are non-topological solitons arising in scalar field theories. Solutions for rotating Q-balls (and the related boson stars) have been shown to exist when the angular momentum is equal to an integer multiple of the Q-ball charge $Q$. Here we consider the possibility of classically long-lived metastable rotating Q-balls with small angular momentum, even for large charge, for all scalar theories that support non-rotating Q-balls. This is relevant for rotating extensions of Q-balls and related solitons such as boson stars as it impacts their cosmological phenomenology.  
}

\preprint{UCI-TR-2022-28}
\arxivnumber{2302.11589}

\flushbottom
\maketitle
\flushbottom

\section{Introduction\label{sec:Intro}}
In a seminal paper~\cite{Coleman:1985ki}, Coleman proved that nontopological solitons called Q-balls exist in scalar field theories where the potential satisfies certain conditions (see also~\cite{Lee:1991ax}). This was extended to supersymmetric models~\cite{Kusenko:1997zq,Enqvist:1997si}, and Q-balls were shown to be good dark matter candidates in both supersymmetric~\cite{Kusenko:1997si,Kusenko:1997vp,Kusenko:2001vu} and nonsupersymmetric~\cite{Krylov:2013qe,Ponton:2019hux,Bai:2019ogh,Bai:2021mzu} theories. Q-balls can be produced by the collapse of an Affleck-Dine condensate~\cite{Multamaki:2002hv,Tsumagari:2009na}, or by the formation of miniclusters in the early universe which subsequently collapse~\cite{Griest:1989cb,Griest:1989bq,Hiramatsu:2010dx,Bai:2022kxq}.

In these discussions, it is usually assumed that  
miniclusters with angular momentum can still collapse into Q-balls. This is a normal assumption for 
collapse to black holes or disks, for which any angular momentum is allowed.
However, for rotating Q-balls~\cite{Volkov:2002aj,Campanelli:2009su,Arodz:2009ye,Shnir:2011gr,Nugaev:2014iva,Loiko:2018mhb} (and the closely related boson stars~\cite{Silveira:1995dh,Kleihaus:2005me,Kleihaus:2007vk,Kleihaus:2011sx,Liebling:2012fv,Davidson:2016uok,Herdeiro:2019mbz,Collodel:2019ohy,Delgado:2020udb,Kling:2020xjj,Dmitriev:2021utv,Gervalle:2022fze,Siemonsen:2023hko}) the dominant paradigm is that the angular momentum must be an integer multiple of the charge $Q$. If this is the case, then the formation of large Q-balls would be drastically different from the usual collapse, since angular momentum would need to be shed in a very precise manner. 

This is, in fact, a puzzling scenario since one might expect that a large classical object like a Q-ball 
 could be given a small angular velocity by adding, for instance, a single particle with nonzero angular momentum about the center of the soliton. This would naively give the final object a small angular momentum. Classically, at least, these are continuous quantities, and one expects to be able
to make the angular velocity and angular momentum arbitrarily small.
Naturally, the angular momentum is quantized in the quantum theory, but even so, one would expect that it should be possible to place a small number of particles in a state of nonzero angular momentum, so that the angular momentum of the Q-ball
does not scale with the total charge and the reported quantization of Q-ball angular momentum is a purely classical effect.

 In this article, we revisit these issues. We begin with a brief history of rotating boson stars, which were analyzed some years before rotating Q-balls. The first published analysis of boson star rotation~\cite{Kobayashi:1994qi} performed a perturbative analysis which indicated that boson stars cannot slowly rotate. However, the authors of this analysis only considered axisymmetric perturbations of the gravitational metric and the scalar field, with no dependence on the azimuthal angle $\varphi$. While axisymmetric perturbations are sufficient to support angular momentum in the gravitational portions of such a system, an axisymmetric scalar field configuration carries no angular momentum. (For a quick review of this fact see Appendix~\ref{s.App}.) Consequently, the axisymmetric ansatz is ill suited for fully exploring the rotation of boson stars. While it may give insight into angular momentum carried by the gravitational field the axisymmetric assumption precludes the scalar field from carrying any angular momentum.

 Indeed, subsequent numerical studies which produced rotating boson stars assume a $\varphi$ dependent scalar field with a profile of the form
 \bea
f(r,\theta,\varphi)=g(r,\theta)\exp(iN\varphi)\,,\label{ansatz}
\eea
for some integer $N$.
Beginning with~\cite{Schunck1996,Schunck:1996he}, these investigation found that the angular momentum $J$ of the boson star satisfies $J=NQ$, where $Q$ is the particle number of the star.\footnote{Similar proportionalities of charge and angular momentum have been found in other soliton systems, see e.g.~Refs.~\cite{Radu:2008pp, Radu:2008ta}.} However, as noted in these first analyses, this quantization of the boson star angular momentum follows directly from the form of the scalar profile assumed in Eq.~\eqref{ansatz}. While the numerical construction of solutions that fit this profile is clearly significant, the fact that those solutions exhibit quantized angular momenta can only be seen as a consequence of the assumed form of the scalar field.

Therefore, the question remains, if a more general ansatz is employed, one that does not begin with the \emph{assumption} $J=NQ$ , can rotating boson stars be found with a different relationship between the angular momentum and particle number? In order to disentangle gravitational effects and focus on the dynamics of the scalar field, in this paper we confine ourselves to the simpler case of rotating Q-balls. So far, all rotating Q-ball solutions have also assumed a scalar field configuration like Eq.~\eqref{ansatz}. Consequently, the solutions have all exhibited $J=NQ$, with the implications for Q-ball production discussed above.

 In this work we analyze the classical equations of motion and consider a more general perturbation around the nonrotating Q-ball than previous analyses. This perturbation allows for small rotations. We find that to leading order in the small angular velocity $\mu$, there is indeed a perturbation which has nonzero angular momentum and is localized near the Q-ball; specifically the profile of the perturbation falls off exponentially far away from the Q-ball. For scalar potentials that produce nonlinearities in the equations of motion, however, higher orders in the perturbation expansion cannot be self-consistently taken to vanish.\footnote{Some boson stars with simple potentials for the scalar field may not require these higher order modes, however, the nonlinear gravitational interactions require a dedicated study.} The generic expectation is that for both Q-balls and boson stars we must include higher order terms, some of which are likely to be oscillatory and fall off only as fast as $1/r$. These contributions suggest that our perturbative ansatz that allows for small rotation also leads to the radiation of energy and angular momentum and is hence an unstable field configuration, not a true solution to the field equations.
 
 The effect of this instability is characterized in order to provide an estimate of the life-time of the localized, rotating perturbation. This is done by calculating the radiated power due to the oscillating modes and comparing that to the energy of the localized, rotating perturbation. For all Q-balls potentials, we find that the decay life-time of this perturbation due to radiation can be very long if the perturbation is small enough. As a consequence, the process of collapsing miniclusters can proceed as usual, at least for sufficiently slow rotation.

After reviewing nonrotating Q-balls in the following section, and setting up our notation, we motivate and present our new more general ansatz for rotating Q-balls in Sec.~\ref{sec:Rot}. The ansatz is based on the expansion of the scalar field in terms of spherical harmonics $Y_{LM}$, with the $Y_{00}$ term taken to be a nonrotating Q-ball. This allows us to smoothly go to the nonrotating Q-ball limit by taking the $L\neq0$ modes to zero. Similarly, we can consider small perturbations away from the known nonrotating solutions by taking these modes to be proportional to $\varepsilon\ll1$.

These higher $L$ modes also involve a new parameter $\mu$, which we show to correspond to the angular velocity of the Q-ball in Appendix~\ref{s.App}. For $\varepsilon\ll1$ and small angular velocities (Sec.~\ref{sec:LinEq}), we find an analytical solution to the equations of motion of the Q-balls to leading order in $\varepsilon$ and $\mu$ for all scalar potentials that lead to Q-balls. Aspects of higher order corrections to this solution are discussed in Sec.~\ref{sec:higher_orders} with several details provided in Appendix~\ref{app:radiation}. These sections outline at what order radiating modes might be sourced by the initial perturbation. We also estimate the lifetime of the metastable rotating state by evaluating the equivalent of the Poynting vector for this scalar field configuration. 
At the classical level, we find that slowly rotating, radiating states can persist for cosmological times. 
We close with a discussion of our results in Sec.~\ref{sec:conclusion}

\section{Non-rotating Q-balls\label{sec:noRot}}
In theories of a complex scalar field $\Phi (t,\vec x)$ with $U(1)$-invariant Lagrangian 
\begin{equation}
\left|\del_\mu\Phi \right|^2-U\left(|\Phi|^2\right)\,,
\end{equation}
 solitonic solutions called Q-balls exist
if the function $U(|\Phi|^2)/|\Phi|^2$ has a minimum at $|\Phi|=\phi_0/\sqrt{2}$ with $0<\phi_0<\infty$, and
\begin{equation}
0\leq\sqrt{\frac{2}{\phi_0^2}U\left(\frac{\phi_0^2}{2}\right)}\equiv \omega_0<m_\phi\,.\label{e.Omega0}
\end{equation}
The Q-ball solutions are found by making the spherically-symmetric ansatz
\beq
\Phi (t,\vec x)=\Phi_0(r = |\vec x|) e^{-i\omega t}\equiv \frac{\phi_0}{\sqrt{2}}f(r) e^{-i\omega t}\,, 
\eeq
and solving the equation of motion
\beq
-\omega^2 \Phi_0- \nabla^2 \Phi_0  +\frac{\del U(|\Phi_0|^2)}{\del \Phi_0^*}=0 \label{nreqn1}
\eeq
for the radial profile $\Phi_0(r)$.
This equation can be shown to have a solution when $\omega_0< \omega<m_\phi$~\cite{Coleman:1985ki}. One way to see that the solutions are localized is to consider the large $r$ (``exterior" to the Q-ball) region. One finds that the profile goes like
\beq
f(r)\sim\frac{c}{r}e^{-r\sqrt{m_\phi^2-\omega^2}}~,\label{e.localProfile}
\eeq
showing that $\omega$ must be less than $m_\phi$. The coefficient $c$ is determined by matching to the interior solution.

It is often preferable to use dimensionless quantities
\begin{align}
\rho&=r\sqrt{m_\phi^2-\omega_0^2}~,& \tau&=t\sqrt{m_\phi^2-\omega_0^2}~,&\nonumber\\
 \overline{\omega}&=\frac{\omega}{\sqrt{m_\phi^2-\omega_0^2}}~,& \overline{U}(|\Phi|^2)&=\frac{U(|\Phi|^2)}{\phi_0^2(m_\phi^2-\omega_0^2)}~,
\end{align}
for analytic and numeric methods to construct these and related solitons~\cite{Heeck:2020bau,Heeck:2021zvk,Heeck:2021gam,Heeck:2021bce,Almumin:2021gax}.
The Q-ball charge $Q$ and energy $E$ are given by  the integrals
\begin{align}
Q &= i\int \dd^{3}x\left(\Phi ^{*}\dot{\Phi} -\dot{\Phi}^{*}\Phi \right) ,\\
E &= \int \dd^{3}x\left[|{\dot {\Phi }}|^{2}+|\nabla \Phi |^{2}+U(|\Phi|^2)\right] .\label{eq:QballEnergy}
\end{align}
For fields that satisfy the equations of motion~\eqref{nreqn1} it is straightforward to show that
\beq
\dd E=\omega\, \dd Q\,,\label{e.dEdQ}
\eeq
which indicates that $\omega$ acts like a chemical potential~\cite{Nugaev:2019vru}. Q-balls that are stable against dissociation and fission have an increasing $Q$ for decreasing $\omega$, with $Q\to \infty$ as $\omega\to \omega_0$~\cite{Coleman:1985ki,PaccettiCorreia:2001wtt}.

\section{Rotating Q-balls \label{sec:Rot}}
Clearly, rotating Q-balls cannot be spherically symmetric, or else the angular momentum $\vec{J}$ would be zero. For a scalar field this is also true of axisymmetric configurations, as shown in Appendix~\ref{s.App}.
In particular, the $z$ component of $\vec{J}$ is given by
\bea
J_z
=-\int \dd^3x\left(\dot{\Phi}\del_\varphi\Phi^\ast+\dot{\Phi}^\ast\del_\varphi\Phi \right) 
\eea
in spherical coordinates. For a nonzero $J_z$, the field $\Phi$ must therefore depend on the azimuthal angle $\varphi$.

Typically~\cite{Volkov:2002aj,Campanelli:2009su,Arodz:2009ye,Shnir:2011gr,Loiko:2018mhb,Silveira:1995dh,Kleihaus:2005me,Kleihaus:2007vk,Kleihaus:2011sx,Liebling:2012fv,Davidson:2016uok,Herdeiro:2019mbz,Collodel:2019ohy,Delgado:2020udb,Gervalle:2022fze}, the profile for the scalar field of a rotating Q-ball or boson star is assumed to take the form
\beq
\Phi(t, r,\theta,\varphi)=f(r,\theta)e^{iN\varphi}e^{-i\omega t}\label{e.BadAn}
\eeq 
with integer $N$, which leads to $J_x = J_y = 0$ and
\bea
J_z
=iN\int \dd^3x\left(\Phi^\ast\dot{\Phi}-\dot{\Phi}^\ast\Phi \right) = NQ\,.
\eea
Note that this follows completely from the form of Eq.~\eqref{e.BadAn}, without any reference to equations of motion. Consequently, any scalar field configuration, soliton or not, of the form given in Eq.~\eqref{e.BadAn} satisfies $J=NQ$. If this field configuration has a large $Q$, then making this ansatz amounts to assuming that the field cannot have small angular momentum.
Numerical solutions for $f(r,\theta)$ have been found, providing evidence for the existence of rotating Q-balls whose angular momentum scales with the charge. 

However, it is not obvious that the $\varphi$ dependence in the scalar field must take the simple form of~\eqref{e.BadAn}.
To explore the possibility of Q-balls with $|J_z|\ll Q$, we must consider a more general ansatz where the profile 
contains different components with different $\varphi$ dependence:
\begin{align}
\Phi
=\frac{\phi_0}{\sqrt{2}}\left[f(r) e^{-i\omega t}
+\sum_{L,M} h_{LM}(r,t)Y_{LM}(\theta,\varphi)\right] ,
\label{ansatz0}
\end{align}
where the $Y_{LM}$ are the usual spherical harmonics; an $L=0$ term can be absorbed into $f(r)$ so the sum begins with $L=1$.
As these ansatze are continuously connected to the non-rotating solution $\Phi_0e^{-i\omega t}$, one expects that perturbations of this form, including those with angular momentum, can be made arbitrarily small, unlike in the ansatz~\eqref{e.BadAn}.
For example if the $h_{LM}$ are small, this could correspond to the introduction of a few particles, possibly including some with angular momentum relative to the center of the soliton, to a nonrotating Q-ball.

To motivate a suitable ansatz for the time dependence, we note that  the ground states of rotating  Q-balls  should have the lowest energy with a fixed charge $Q$ and angular momentum $J_z$.
These are found by introducing two Lagrange multipliers $\omega$ and $\mu$, and minimizing the functional
\begin{align}
E_{\omega,\mu} &= E+\omega \left[Q-i\int \dd^{3}x(\Phi ^{*}\dot{\Phi} -\Phi \dot{\Phi} ^{*})\right]+\mu\left[J_z+\int \dd^3x\left(\dot{\Phi}^* \del_\varphi\Phi+\dot{\Phi} \del_\varphi\Phi^*\right)\right]
\\
&=\omega Q+\mu J_z+\int \dd^{3}x\left[|{\dot {\Phi }}+i\omega \Phi+\mu\del_\varphi\Phi |^{2}+|\nabla \Phi |^{2}+ U - \left|i\omega\Phi+\mu\partial_\varphi\Phi \right|^2\right] ~.
\end{align}
Minimizing the first term in the integral leads to
\begin{align}
\dot{\Phi }+i\omega \Phi+\mu\del_\varphi\Phi=0\,,\label{e.Phidot}
\end{align}
which ensures  that charge, angular momentum about the $z$-axis, and energy are time independent.
Equation~\eqref{e.Phidot} also implies that
\bea
h_{LM}(r,t)=h_{LM}(r)e^{-i(\omega+M\mu) t}  \,.
\label{ansatz1}
\eea

Applying Eq.~\eqref{e.Phidot} allows us to write the energy~\eqref{eq:QballEnergy} as
\beq
E=\omega Q+\mu J_z-\mathcal{L}~,\label{e.EnergySmarr}
\eeq
where the Lagrangian $\mathcal{L}$ is given by
\beq
\mathcal{L}=\int \dd^3x\left[\omega^2|\Phi|^2+\mu^2\left|\partial_\varphi\Phi\right|^2+i\omega \mu  \left(\Phi\partial_\varphi\Phi^\ast-\Phi^\ast\partial_\varphi\Phi \right)-\left|\nabla\Phi\right|^2-U\right]~.
\eeq
For fields that satisfy the equations of motion, one can show that
\beq
\frac{\dd \mathcal{L}}{\dd \omega}=Q~, \ \ \ \frac{\dd \mathcal{L}}{\dd\mu}=J_z~.
\eeq
Straightforward calculations produce the relations
\begin{align}
\frac{\dd E}{\dd\omega}&=Q+\omega\frac{\dd Q}{\dd\omega}+\mu\frac{\dd J_z}{\dd\omega}-\frac{\dd \mathcal{L}}{\dd\omega}\nonumber\\
&=\omega\frac{\dd Q}{\dd\omega}+\mu\frac{\dd J_z}{\dd\omega}\,,\\
\frac{\dd E}{\dd \mu}&=\omega\frac{\dd Q}{\dd\mu}+J_z+\mu\frac{\dd J_z}{\dd \mu}-\frac{\dd \mathcal{L}}{\dd \mu}\nonumber\\
&=\omega\frac{\dd Q}{\dd\mu}+\mu\frac{\dd J_z}{\dd\mu}\,.
\end{align}
In other words, we see that Eq.~\eqref{e.dEdQ} is generalized to
\beq
\dd E=\omega\, \dd Q+\mu\, \dd J_z\,.
\eeq
Since $\mu$ is conjugate to $J_z$, it should be related to 
the angular velocity of the soliton about the $z$-axis. This is also seen in the form of our ansatz~\eqref{ansatz1}, which depends on the combination $\varphi-\mu t$.
Finally, it is shown in the appendix~\ref{s.App} that 
\beq
\frac{\dd J_x}{\dd t}=-\mu J_y\,, &&
\qquad \frac{\dd J_y}{\dd t}=\mu J_x\,,\label{e.Euler}
\eeq
which correspond to the Euler equations of a rotating rigid body in the absence of external torques, when $\mu$ is taken to be the angular velocity about the $z$-axis.

Notice that the well-known ansatz from Eq.~\eqref{e.BadAn} is rather special: since $J_z$ and $Q$ are not independent in this ansatz, $\mu$ and $\omega$ are not independent either and always show up as a fixed linear combination $\omega + N \mu$. Existing numerical solutions do not reference $\mu$, even though they find solitons with nonzero angular velocity. These works effectively absorb $\mu$ into their working definition of $\omega$.
Our above formalism is useful precisely for the opposite scenario in which $J_z$ and $Q$ can be varied independently.

\section{A Small \texorpdfstring{$J_z$}{Lz} Solution \label{sec:LinEq}}
As we construct the equations that determine the ansatz \eqref{ansatz0}, in the limit of small $J_z$,
we use the spherical harmonic convention 
$
Y_{L,M}^\ast=(-1)^MY_{L,-M} \,.
$
This allows us to write
\begin{align}
\Phi^\ast=\frac{\phi_0}{\sqrt{2}}\left[fe^{i\omega t}
+\sum_{L,M} h^\ast_{L,-M}(-1)^MY_{LM}e^{i(\omega-M\mu) t}\right]
\end{align}
as a sum over the same spherical harmonics as in~\eqref{ansatz1}.  The time dependence is absent from the equations of motion and the conserved quantities, making it useful to define the fields
\beq
h^{\pm}_{LM}\equiv h_{LM}\pm(-1)^Mh^\ast_{L,-M}~,
\eeq
where
\beq
h^{\pm\ast}_{LM}=\pm(-1)^Mh^\pm_{L,-M}~.\label{e.Hcondition}
\eeq

We also express the angular momentum and charge of the soliton in terms of these fields. The above relations allow us to do so while only summing over $M\geq0$. First, we find 
\beq
J_z=\frac{\phi_0^2}{2}\sum_{L,M>0}M\int dr\,r^2\left[\mu M\left(|h_{LM}^+|^2+|h_{LM}^-|^2 \right)+\omega\left(h^{+\ast}_{LM}h^-_{LM}+h^{-\ast}_{LM}h^+_{LM} \right) \right]~.
\eeq 
Note that we expect the angular momentum to vanish for $\mu=0$ when $\omega>0$. This is guaranteed to occur if either $h^+_{LM}$ or $h^-_{LM}$ goes to zero as $\mu\to0$. We also find the charge
\begin{align}
Q=&4\pi\omega\phi_0^2\int dr\,r^2f^2+\frac{\phi_0^2}{2}\sum_{L,M>0}\int\! dr\,r^2\left[\omega\left(|h_{LM}^+|^2+|h_{LM}^-|^2 \right)+\mu M\left(h^{+\ast}_{LM}h^-_{LM}+h^{-\ast}_{LM}h^+_{LM} \right) \right]~\nonumber\\
&+\frac{\omega\phi_0^2}{4}\sum_{L}\int dr\,r^2\left(|h_{L0}^+|^2+|h_{L0}^-|^2 \right)~.
\end{align}
When $\mu\to0$ the corrections to $Q$ from this perturbation need not vanish. Such a scenario might correspond to the introduction of additional scalar field that does not carry angular momentum.

In general,
the $h^\pm_{LM}$ of different $(L,M)$ are all coupled since the equations of motion are nonlinear. However, for small perturbations $h_{LM}\to\epsilon\, h_{LM}$ with $\epsilon\ll1$, the $\epsilon^1$ order functions decouple:
\begin{align}
0&= \partial_{\rho}^2 h^+_{LM}+\frac{2}{\rho}\partial_{\rho} h^+_{LM}-\frac{L(L+1)}{\rho^2}h^+_{LM}+\left(\overline{\omega}^2+M^2\overline{\mu}^2\right)  h^+_{LM}+2\overline{\omega} M\overline{\mu} \,h^-_{LM}\nonumber\\
&\quad -h^+_{LM}\phi_0^2\left.\frac{\del \overline{U}}{\del  (\Phi\Phi^*)}\right|_{\Phi=\Phi_0}-h^+_{LM}f^2\phi_0^4\left. \frac{\del^2\overline{U}}{\del  (\Phi\Phi^*)^2}\right|_{\Phi=\Phi_0} , \label{e.hplus}\\
0&=\partial_{\rho}^2 h^-_{LM}+\frac{2}{\rho}\partial_{\rho} h^-_{LM}-\frac{L(L+1)}{\rho^2}h^-_{LM}+\left(\overline{\omega}^2+M^2\overline{\mu}^2\right)  h^-_{LM}+2\overline{\omega} M\overline{\mu} \,h^+_{LM}\nonumber\\
&\quad-h^-_{LM}\phi_0^2\left. \frac{\del \overline{U}}{\del  (\Phi\Phi^*)}\right|_{\Phi=\Phi_0} ,\label{e.hminus}
\end{align}
where 
\beq
\overline{\mu}=\frac{\mu}{\sqrt{m_\phi^2-\omega_0^2}}~.
\eeq
For each $(L,M)$, these are two real coupled differential equations. The leading order contributions to the energy are
\begin{align}
E=&\frac{\phi_0^2}{2}\int dr r^2\left[\left(\partial_rf\right)^2+\omega^2f^2+U\left(f^2\right) \right]\nonumber\\
&+\epsilon^2\frac{\phi_0^2}{4}\sum_{L,M>0}\int dr r^2\left\{\left|\partial_r h^+_{LM} \right|^2+\left|\partial_r h^-_{LM} \right|^2+2M\omega\mu \left(h^{+\ast}_{LM}h^-_{LM}+h^{-\ast}_{LM}h^+_{LM} \right) \phantom{\frac{U}{\phi}}\right.\nonumber\\
&+\left[\mu^2 M^2+\omega^2+\frac{L(L+1)}{r^2} +\left.\frac{\del U}{ \del  (\Phi\Phi^*)}\right|_{\Phi=\Phi_0}\right]\left(|h_{LM}^+|^2+|h_{LM}^-|^2 \right)\nonumber\\
&\left.+2\Phi^2_0\left|h^+_{LM}\right|^2\left.\frac{\del^2U}{\del  (\Phi\Phi^*)^2}\right|_{\Phi=\Phi_0}\right\}
+\epsilon^2\frac{\phi_0^2}{8}\sum_{L}\int dr r^2\left\{\left|\partial_r h^+_{L0} \right|^2+\left|\partial_r h^-_{L0} \right|^2\phantom{\frac{U^2}{\Phi^2}} \right.\\
&\left.+\left[\omega^2+\frac{L(L+1)}{r^2} +\left.{\del U\over \del  (\Phi\Phi^*)}\right|_{\Phi=\Phi_0}\right]\left(|h_{L0}^+|^2+|h_{L0}^-|^2 \right)+2\Phi^2_0\left|h^+_{L0}\right|^2\left.{\del^2U\over \del  (\Phi\Phi^*)^2}\right|_{\Phi=\Phi_0}\right\}~.\nonumber
\end{align}
We emphasize that at this point the perturbation parameter $\epsilon$ used here captures both corrections with angular momentum and those without.

 To obtain solutions that correspond to small angular momentum, we further expand each $h^\pm_{LM}$ in a power series in $\overline{\mu}$. We note that $\overline{\mu}\to -\overline{\mu}$ leaves the equations of motion invariant if combined with $h^\pm_{LM}\to \pm h^\pm_{LM}$. The expansion therefore has the form
\begin{align}
h^+_{LM}&=h^{+(0)}_{LM}+\overline{\mu}^2 h^{+(2)}_{LM}+\ldots
\\
h^-_{LM}&=\overline{\mu}\, h^{-(1)}_{LM}+\overline{\mu}^3 h^{-(3)}_{LM}+\ldots~.
\end{align}
In particular, we see that $h^+_{LM}$ can be nonzero even when $\overline{\mu}\to0$.
By a nearly identical argument we see that $h^\pm_{L,-M}=\pm h^\pm_{LM}$. We set all $h_{L0}=0$ since these modes do not contribute to $J_z$ even for nonzero $\overline{\mu}$.

To zeroth order in $\overline{\mu}$, the $h^+_{LM}$ equation is
\begin{align}
0=\,& \partial_{\rho}^2 h^{+(0)}_{LM}+\frac{2}{\rho}\partial_{\rho} h^{+(0)}_{LM}-\frac{L(L+1)}{\rho^2}h^{+(0)}_{LM}
+\overline{\omega}^2 h^{+(0)}_{LM} \nonumber\\
& -h^{+(0)}_{LM}\left[ \phi_0^2\frac{\del \overline{U}}{\del  (\Phi\Phi^*)}
+f^2\phi_0^4\frac{\del^2\overline{U}}{\del  (\Phi\Phi^*)^2}\right]_{\Phi=\Phi_0} .\label{e.hPlusMu0}
\end{align}
 To find a solution, we take the $\rho$ derivative of Eq.~\eqref{nreqn1} and find 
\begin{align}
0 &= -\overline{\omega}^2 \partial_{\rho}f-  \partial_{\rho}^3 f-\frac{2}{\rho}\partial_{\rho}^2 f
+\frac{2}{\rho^2}\partial_{\rho} f+\partial_{\rho} f \phi_0^2\frac{\del \overline{U}}{\del (\Phi_0\Phi_0^*)}+\phi_0^4f^2\partial_{\rho} f \frac{\del^2 \overline{U}}{\del (\Phi_0\Phi_0^*)^2}\,.
\end{align}
This shows that there is an {\it exact} solution of Eq.~\eqref{e.hPlusMu0} for $L=1$:
\bea
 h^{+(0)}_{1,\pm1}=c_1\partial_{\rho} f\,,
\eea
where the constant $c_1$ must be purely imaginary to satisfy Eq.~\eqref{e.Hcondition}. As the magnitude is arbitrary at this order in $\epsilon$, we simply take $c_1=$i. 

\begin{figure}[tb]	
\centering
\includegraphics[width=0.49\textwidth]{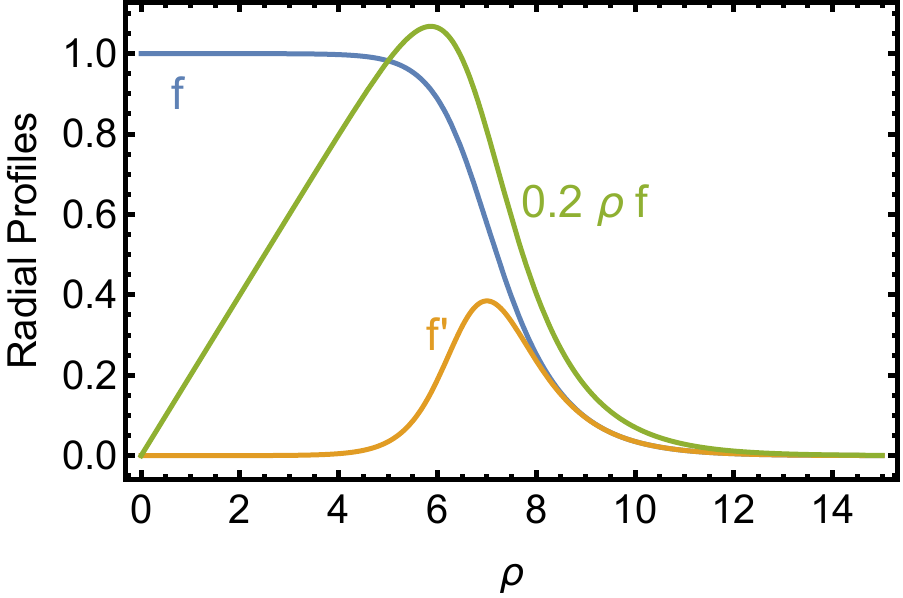}  \includegraphics[width=0.49\textwidth]{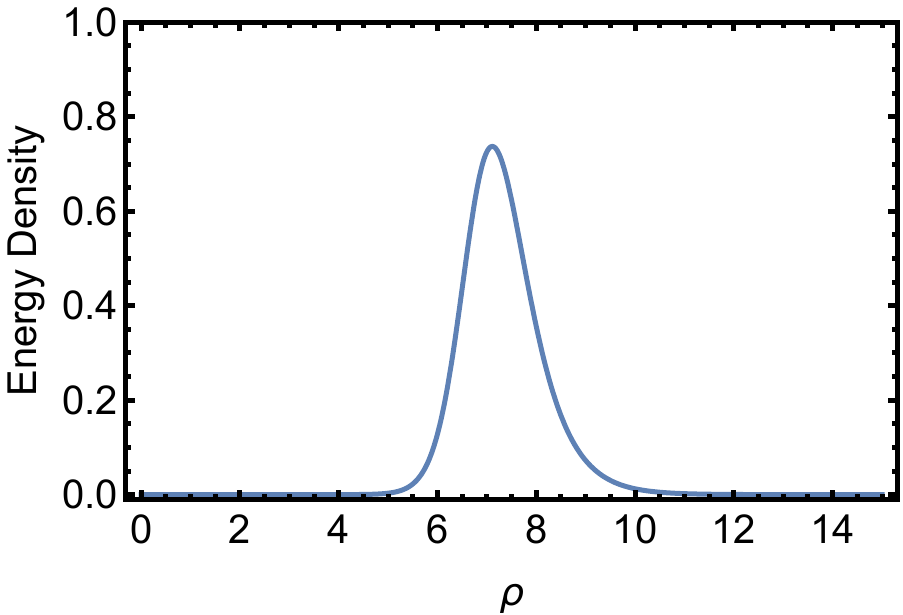}  
	\caption{(Left) Illustration of the radial profiles $f$, $h^{+(0)}_{1,-1}$, and $h^{-(1)}_{1,-1}$ and (right) the $\epsilon^2\overline{\mu}^0$ correction to the energy density for the sextic potential of Ref.~\cite{Heeck:2020bau}. The normalization of the perturbations, including the correction to the energy density, is has been adjust to make them fit easily on the same figure.}
	\label{fig:profiles}
\end{figure}

To first order in $\bar\mu$, the equation for $h^-_{11}$ is
\begin{align}
\partial_{\rho}^2 h^{-(1)}_{1,\pm1}+\frac{2}{\rho}\partial_{\rho} h^{-(1)}_{1,\pm1}-\frac{2}{\rho^2}h^{-(1)}_{1,\pm1}+\overline{\omega}^2 h^{-(1)}_{1,\pm1}
-h^{-(1)}_{1,\pm1}\phi_0^2\left. \frac{\del \overline{U}}{\del  (\Phi\Phi^*)}\right|_{\Phi=\Phi_0}=\mp\, 2\text{i}\omega \partial_{\rho}f\,.
\end{align}
One can verify, using equation~\eqref{nreqn1}, that this is solved by
\beq
h^{-(1)}_{1,\pm1}=\mp\, \text{i}\overline{\omega}\, \rho f~.
\eeq
Thus for $L=1$ we have found a solution up to order $\overline{\mu}$
\begin{align}
 h^{+}_{1,\pm1}=\text{i}\partial_{\rho} f\,, &&  h^{-}_{1,\pm1}=\mp\text{i}\overline{\mu}\,\overline{\omega}\, \rho f \,.
\end{align}
We illustrate these profiles in the left panel of Fig.~\ref{fig:profiles} for a sextic potential, but emphasize that the solution holds for any potential that supports Q-balls. For this configuration the $\epsilon^2\overline{\mu}^0$ contribution to the energy density, as calculated in what follows, has the form given in the right panel of Fig.~\ref{fig:profiles}. Note that because the energy is independent of the direction of rotation the first $\overline{\mu}$ dependence comes at order $\overline{\mu}^2$. To obtain the complete correction to this order we would also need the $\overline{\mu}^2$ contribution to $h^{+}_{1,\pm1}$.

Now, the unperturbed solution (with $J_z=0$) has particle number and energy~\cite{Heeck:2020bau} given by 
\begin{align}
Q_0 &= \frac{4\pi\overline{\omega}\phi_0^2}{m_\phi^2-\omega_0^2}\int \dd \rho\,\rho^2 f^2 \,,
\\
E_0&=\omega Q_0+\frac{4\pi\phi_0^2}{3\sqrt{m_\phi^2-\omega_0^2}}\int  \dd \rho\,\rho^2\left(\partial_{\rho}f\right)^2 . 
\end{align}
The first term in the energy scales like the volume of the Q-ball, while the second scales like the surface area and so is typically subleading for large Q-balls. 

Using the pertubative solution given above we find that the pertubation contributes an angular momentum that is positive for $\overline{\mu}>0$ and to leading order is
\begin{align}
\Delta J_z=& \frac{\phi_0^2\overline{\mu}\epsilon^2}{m_\phi^2-\omega_0^2}
\int \dd\rho\, \rho^2\left[\left(\partial_{\rho}f\right)^2+3\overline{\omega}^2f^2\right]+\mathcal{O}(\epsilon^2\overline{\mu}^2) \nonumber\\
=&\overline{\mu}\epsilon^2\frac{3}{4\pi}\frac{E_0}{\sqrt{m_\phi^2-\omega_0^2}}+\mathcal{O}(\epsilon^2\overline{\mu}^3)\,.
\end{align}
with $J_{x,y} =0$.
We also find that the charge is shifted from $Q_0$ by a positive amount 
\begin{align}
&\Delta Q=\epsilon^2\frac{3}{4\pi} \frac{E_0-\omega Q_0}{\sqrt{m_\phi^2-\omega_0^2}}+\mathcal{O}(\epsilon^2\overline{\mu}^2)
\end{align}
and the energy is shifted from the nonrotating value by
\begin{align}
\Delta E &= \omega\Delta Q+\mathcal{O}(\epsilon^2\overline{\mu}^2)\,.
\end{align}
Notice that, unlike the angular momentum, the leading $\overline{\mu}$ corrections to the charge and energy go like $\overline{\mu}^2$ so our linear in rotation solution does not capture all the possible energy corrections to leading order. 

We also compare the energy of the localized solution to one in which the added charge $\Delta Q$ is spread at infinity with zero kinetic energy and hence has $\Delta E=m_\phi\Delta Q$. This delocalized solution has higher energy as long as $\omega<m_\phi$ (which is always true for a Q-ball). This indicates that when the higher order terms can be consistently set to zero that the rotating perturbative solution is stable against dissociation.

\section{Higher Orders in $\epsilon$}
\label{sec:higher_orders}

At linear order in $\epsilon$ the system of equations~\eqref{e.hplus} and~\eqref{e.hminus} is solvable, as illustrated above, as a perturbation series in $\overline{\mu}$. Issues arise at higher order in $\epsilon$, where the non-linearities in the Q-ball potential couple the different $(L,M)$ modes and enforce an infinite set of non-vanishing $h_{LM}$. 

Exterior to the Q-ball, at large $r$, the relevant differential equation of some higher-$\epsilon$ mode takes the form
\begin{align}
\partial_{r}^2 h_{LM}+\frac{2}{r}\partial_{r} h_{LM}-\frac{L(L+1)}{r^2}h_{LM}+\left(\omega^2 - m_\phi^2+M^2\mu^2\right)  h_{LM} \approx 0\,,
\label{eq:radiation}
\end{align}
where exponentially suppressed terms have been dropped on the right hand side.
As shown in App.~\ref{app:radiation}, for almost all Q-balls consistency requires infinitely many $(L,M)$ modes. At large $M$, the expression $\omega^2- m_\phi^2+M^2\mu^2$ becomes positive and no longer admits localized solutions, but rather corresponds to radiation modes that indicate an instability.

In this case the solutions to Eq.~\eqref{eq:radiation} are spherical Bessel functions. However, we are primarily interested in their large-$r$ form
\beq
h_{LM}\sim \frac{1}{r\sqrt{\omega^2-m_\phi^2+M^2\mu^2}}\cos\left(r\sqrt{\omega^2-m_\phi^2+M^2\mu^2}-L\frac{\pi}{2} \right)~.
\eeq
As shown below, because these field fall off like $1/r$ they lead to nonzero radiated power.

While this conclusion is fairly straightforward, we highlight a subtle feature: the right-hand side in Eq.~\eqref{eq:radiation} is of course not zero, but rather exponentially suppressed by the lower-$\epsilon$ source terms. The resulting \emph{inhomogeneous} differential equation could conceivably have a localized solution, allowing us to drop the oscillating solution to the \emph{homogeneous} equation.
The answer to this question lies beyond the scope of this article, in the following we work under the conservative assumption that the classical slowly-rotating Q-ball solution is indeed unstable.

The lifetime of these solitons is impossible to calculate exactly without calculating the perturbative series up to the order nonlocalized functions are sourced. 
To estimate the lifetime of the rotating perturbation to the spherical Q-ball solution we separate the perturbation into two parts: the localized field and the radiating field. This radiating field carries angular momentum and energy away from the localized field configuration. Similar to the Poynting vector analysis in electromagnetism, the radiated power carried by the nonlocalized modes at an instant in time is determined by
\beq
\mathcal{P}=\left.\int\dd \theta\,\dd\varphi\, r^2\sin\theta\, T^{0r} \right|_{r\to\infty}~,
\eeq
where the energy-momentum tensor is 
\beq
T_{\mu\nu}=\partial_\mu(\Phi^\ast)\partial_\nu\Phi+\partial_\nu(\Phi^\ast)\partial_\mu\Phi-\eta_{\mu\nu}\left[\partial_\alpha(\Phi^\ast)\partial^\alpha\Phi-U(\Phi^\ast\Phi) \right]~.
\eeq
This leads to
\beq
\mathcal{P}=\left.\int\dd \theta\,\dd\varphi\, r^2\sin\theta\left[\text{i} \omega\left(\Phi\partial_r\Phi^\ast-\Phi^\ast\partial_r\Phi \right)+\mu\left(\partial_\varphi(\Phi^\ast)\partial_r\Phi+\partial_r(\Phi^\ast)\partial_\varphi\Phi \right) \right]\right|_{r\to\infty}~.
\eeq
The integration is of a quadratic function of the fields over the surface at infinity. 

The scalar field determined in the previous section has the form
\beq
\Phi=\text{localized}+\mathcal{O}(\epsilon^2)+\mathcal{O}(\epsilon\overline{\mu}^2)~.
\eeq
In other words, we have only established that the full perturbative solution is localized to a low order in our expansion parameters $\epsilon$ and $\overline{\mu}$. These localized fields do not contribute to any radiated power. Thus, we conclude that the radiated power, which would need two powers of functions that fall off like $1/r$, goes like
\beq
\mathcal{P}=0+\mathcal{O}(\epsilon^4)+\mathcal{O}(\epsilon^2\overline{\mu}^4)+\mathcal{O}(\epsilon^3\overline{\mu}^2)~.
\eeq
Here $\epsilon,\,\overline{\mu}\ll1$, so the radiated power is small. This allows us to estimate the total time to radiate away all the energy $\Delta E$ and angular momentum of the rotating state as
\beq
\Delta t\sim\frac{\Delta E}{\mathcal{P}}\sim \frac{1}{\mathcal{O}(\epsilon^2)+\mathcal{O}(\overline{\mu}^4)+\mathcal{O}(\epsilon\overline{\mu}^2)},
\eeq
where we have taken the increase of energy from the spinning state to go like $\epsilon^2$, as explicitly calculated in the previous section.\footnote{This follows from the largest shift to the energy coming from the leading order $\epsilon$ correction rather than the higher order terms that might contribute to radiation.} For small $\epsilon$ and $\overline{\mu}$ this time can be arbitrarily long, rendering these field configurations classically metastable. 

The eventual decay may be understood as follows: in  a classical universe, a rotating Q-ball can reduce its angular momentum by emitting an arbitrarily small amount of charge, which can carry an arbitrarily large angular momentum if it moves far away from the Q-ball with a small angular velocity. If the emitted charge can be made arbitrarily small, the binding energy is also small, and this emission is energetically favorable. While this picture is plausible, it does not distinguish between the known solutions with quantized angular momenta and the small angular momenta configurations we are exploring. We leave the full understanding of rotating Q-ball stability to future work.

\section{Conclusion}
\label{sec:conclusion}

In this work we examine for the first time slowly rotating Q-balls.\footnote{Previous perturbative analyses of \textit{boson stars} were not sufficiently general to allow for scalar fields with angular momentum.}  We construct a more general ansatz that allows us to consider perturbative extensions of known Q-ball solutions whose angular momentum need not satisfy $J=NQ$. A localized, leading-order perturbative correction is found which applies to all theories that give rise to Q-balls. 
We further show that this perturbative solution has lower energy than the unperturbed Q-ball surrounded by a free scalar field at infinity and hence is stable against dissociation.

In general, Q-ball potentials produce non-linearities in the equation of motion which require higher order modes to be nonzero. We have shown that this suggests the existence of a non-local, hence radiating, portion of the scalar field. For small rotation, however, these radiation modes are only required to appear at subleading order in the expansion parameter. Therefore, though the localized perturbation is only an approximate solution to the equations of motion, the approximation can be quite close to a true solution. We estimate the life-time of the localized perturbation from the radiated power contained in the non-localized part of the field. We find that small angular momentum solitons are classically metastable in the small angular velocity limit with lifetimes that can be relevant to cosmological studies.

 Therefore, we have meaningfully expanded the possibilities for rotating Q-balls. This is not simply a formal question; there are important phenomenological consequences. Specifically, the existence of this new class of long-lived, slowly rotating Q-ball perturbations enhances the validity of standard calculations of the Q-ball relic density; without these rotating states, only Q-balls with specific values of the angular momentum would be produced, and the relic density would presumably be much smaller. A full analysis requires further investigation into the dynamics of rotating Q-balls and their localized perturbations.

There are many other questions regarding rotating Q-balls. 
We focused on small angular velocities in order to explore the dynamics of rotating 
Q-balls in this limit. It would be interesting to have an improved characterization of 
these solutions for arbitrary angular momentum, perhaps using the methods 
of~\cite{Kling:2017mif,Kling:2017hjm}. Extending these
solutions to boson stars, oscillons, axion stars, and other solitons in the literature would also be very interesting. Also, studying rotating Q-balls in a full quantum mechanical theory would be worth investigating as this might significantly affect their stability. We hope to return to these questions in future work.

\section*{Acknowledgements}
We are grateful to Ruth Gregory, Eric Hirschmann, and Nicol\'as Yunes for helpful discussions. The work of Y.A. and A.R. is supported by the National Science Foundation Grant No.~PHY-1915005. The research of Y.A.~was supported
by Kuwait University. J.H. is supported in part by the National Science Foundation under Grant No. PHY-2210428. C.B.V. is supported in part by the National Science Foundation under Grant No. PHY-2210067.

\appendix
\section{Angular Velocity\label{s.App}}
In this appendix we demonstrate in what sense $\mu$ can be interpreted as the angular velocity of the soliton. The angular momentum $M$ is defined in terms of the energy momentum tensor $T$ by
\beq
M^{ij}=\int d^3x\left(x^iT^{j0}-x^jT^{i0} \right)~,
\eeq
where in Cartesian coordinates $J_x=M^{23}$, $J_y=-M^{13}$, and $J_z=M^{12}$. This leads to
\begin{align}
J_x=&\int \dd^3x\left[\dot{\Phi}\left( \sin\varphi\partial_\theta+\frac{\cos\varphi}{\tan\theta}\partial_\varphi  \right)\Phi^\ast+\dot{\Phi}^\ast\left( \sin\varphi\partial_\theta+\frac{\cos\varphi}{\tan\theta}\partial_\varphi  \right)\Phi\right]~,\\
J_y=&\int \dd^3x\left[\dot{\Phi}\left( -\cos\varphi\partial_\theta+\frac{\sin\varphi}{\tan\theta}\partial_\varphi  \right)\Phi^\ast+\dot{\Phi}^\ast\left( -\cos\varphi\partial_\theta+\frac{\sin\varphi}{\tan\theta}\partial_\varphi  \right)\Phi\right]~,\\
J_z=&-\int \dd^3x\left[\dot{\Phi}\partial_\varphi \Phi^\ast+\dot{\Phi}^\ast\partial_\varphi\Phi\right]~.
\end{align}
Note that from these results we quickly see that for an axisymmetric field $\Phi(r,\theta)$ we have
\begin{align}
J_x=&\int \dd r\dd\theta r^2\sin\theta\left(\dot{\Phi}\partial_\theta\Phi^\ast+\dot{\Phi}^\ast\partial_\theta\Phi \right)\int_0^{2\pi}d\varphi \sin\varphi=0\\
J_y=&-\int \dd r\dd\theta r^2\sin\theta\left(\dot{\Phi}\partial_\theta\Phi^\ast+\dot{\Phi}^\ast\partial_\theta\Phi \right)\int_0^{2\pi}d\varphi \cos\varphi=0\\
J_z=&0~.
\end{align}
This shows that an axisymmetric scalar field configuration has zero angular momentum.

Using the relation given in Eq.~\eqref{e.Phidot} and the above definitions of the components of angular momentum we find
\begin{align}
\frac{\dd J_x}{\dd t}=&\int\dd^3x\left\{\left(-i\omega\dot{\Phi}-\mu\partial_\varphi\dot{\Phi} \right)\left( \sin\varphi\partial_\theta+\frac{\cos\varphi}{\tan\theta}\partial_\varphi  \right)\Phi^\ast\right.\nonumber\\
&+\dot{\Phi}\left( \sin\varphi\partial_\theta+\frac{\cos\varphi}{\tan\theta}\partial_\varphi  \right)\left(i\omega\Phi^\ast-\mu\partial_\varphi\Phi^\ast \right)\nonumber\\
&+\left(i\omega\dot{\Phi}^\ast-\mu\partial_\varphi\dot{\Phi}^\ast \right)\left( \sin\varphi\partial_\theta+\frac{\cos\varphi}{\tan\theta}\partial_\varphi  \right)\Phi\nonumber\\
&\left.+\dot{\Phi}^\ast\left( \sin\varphi\partial_\theta+\frac{\cos\varphi}{\tan\theta}\partial_\varphi  \right)\left(-i\omega\Phi-\mu\partial_\varphi\Phi \right)\right\}\nonumber\\
=&\mu\int\dd^3x\left\{\dot{\Phi}\left( \cos\varphi\partial_\theta-\frac{\sin\varphi}{\tan\theta}\partial_\varphi  \right)\Phi^\ast+\dot{\Phi}^\ast\left( \cos\varphi\partial_\theta-\frac{\sin\varphi}{\tan\theta}\partial_\varphi  \right)\Phi \right\}\nonumber\\
=&-\mu J_y~,
\end{align}
where in the second equality we have integrated by parts to change $\partial_\varphi\dot{\Phi}$ to $\dot{\Phi}$. By a similar calculation we find
\begin{align}
\frac{\dd J_y}{\dd t}=&\mu J_x~,\\
    \frac{\dd J_z}{\dd t}=&-\int \dd^3x\partial_\varphi\left[|\dot{\Phi}|^2-\omega^2|\Phi|^2+\mu^2\left|\partial_\varphi\Phi\right|^2 \right]=0~.
\end{align}
This confirms the results in Eq.~\eqref{e.Euler} that $\mu$ appears as the angular velocity about the $z$-axis in the Euler equations for rigid rotation.

\section{Radiation Modes\label{app:radiation}}
In this appendix we give a more detailed discussion of the non-linearities that couple different $(L,M)$ modes, as mentioned in Sec.~\ref{sec:higher_orders}.
At higher orders in $\epsilon$, nonlinear terms induce higher modes with larger values of $M$ of the form \eqref{ansatz1}. This can be seen more clearly by considering the following alternate expansion of $\Phi$
\beq
\Phi=e^{\text{i}\omega t}\sum_{n=-\infty}^\infty c_n(r,\theta)e^{\text{i}n(\varphi-\mu t)}~,
\eeq
where the time dependence has been chosen so that Eq.~\eqref{e.Phidot} is satisfied. Within the Lagrangian we can consider how the $c_n$ are linked by the potential through non-linear terms like
\beq
\int \dd\varphi\left(|\Phi|^2\right)^N=\int\dd\varphi\prod_{i=1}^N\sum_{n_i=\infty}^{\infty}\sum_{m_i=-\infty}^{\infty} c_{n_i}c_{m_i}^\ast e^{\text{i}\varphi(n_i-m_i)}e^{\text{i}\mu t(m_i-n_i)} ~.
\eeq
When integrated over $\varphi$, terms containing
\beq
\exp\left\{\text{i}\varphi\left(n_1-m_1+\ldots+n_N-m_N \right) \right\}~
\eeq
vanish unless
\beq
n_1-m_1+\ldots+n_N-m_N=0~.
\eeq
This ensures that the potential energy is time independent, but also generally couples low $n$ modes to arbitrarily high $n$. An exceptional case is having only one nonzero $c_n$ where
\beq
\int \dd\varphi\left(|\Phi|^2\right)^N=2\pi\left(|c_n|^2\right)^N~.
\eeq
In general the $n$th equation includes a source term from the potential
\beq
N \prod_{i=1}^{N-1}\sum_{n_i,m_i} c_{n_i}c_{m_i}^\ast c_{n+n_1-m_1+\ldots +n_{i}-m_{i}}~.
\eeq
Consider the simple case of $N=2$ and suppose that two $c_n$ are nonzero: $c_a$ and $c_b$. The equation for some mode $c_n$ includes the source term
\beq
2\sum_{s,t}c_sc^\ast_tc_{n+s-t}~.
\eeq
This contributes to the $n=a$ and $n=b$ equations, but also sources the $n=2a-b$ and $n=2b-a$ modes. Therefore, we must take these modes to be nonzero, which in turn sources more modes, leading to $n$ of larger and larger magnitude. One can show that the modes $n=a+N(b-a)$ and $n=b+N(a-b)$ are sourced for all integers $N$. 

We can now determine at what order in $\epsilon$ higher $M$ models are sourced. In general, this depends on the nature of the initial perturbative solution. Let us use, like the solution obtained in Sec.~\ref{sec:LinEq}, a function which at order $\epsilon^0$ is the nonrotating $M=0$ ($c_0$) solution and the $\epsilon^1$ solution has $L=1$ with $M=\pm1$ ($c_{\pm1}$). This solution is exact to this order in perturbation theory.

The next $M$s that are sourced are $M=\pm2$ and these require the contributions of two $M=\pm1$ modes, so they enter at $\epsilon^2$. This pattern persists at each order in $\epsilon$. Additional $M$ terms require an $M-1$ term coupled to a $M=\pm1$ term through the potential. Thus, in general the $\pm M$ modes enter at order $\epsilon^M$.

\bibliographystyle{JHEP}
\bibliography{BIB}{}

\providecommand{\href}[2]{#2}\begingroup\raggedright\begin{thebibliography}{10}

\bibitem{Coleman:1985ki}
S.~R. Coleman, \emph{{Q Balls}},
  \href{https://doi.org/10.1016/0550-3213(85)90286-X}{\emph{Nucl. Phys.}
  {\bfseries B262} (1985) 263}.

\bibitem{Lee:1991ax}
T.~D. Lee and Y.~Pang, \emph{{Nontopological solitons}},
  \href{https://doi.org/10.1016/0370-1573(92)90064-7}{\emph{Phys. Rept.}
  {\bfseries 221} (1992) 251}.

\bibitem{Kusenko:1997zq}
A.~Kusenko, \emph{{Solitons in the supersymmetric extensions of the standard
  model}}, \href{https://doi.org/10.1016/S0370-2693(97)00584-4}{\emph{Phys.
  Lett. B} {\bfseries 405} (1997) 108}
  [\href{https://arxiv.org/abs/hep-ph/9704273}{{\ttfamily hep-ph/9704273}}].

\bibitem{Enqvist:1997si}
K.~Enqvist and J.~McDonald, \emph{{Q balls and baryogenesis in the MSSM}},
  \href{https://doi.org/10.1016/S0370-2693(98)00271-8}{\emph{Phys. Lett. B}
  {\bfseries 425} (1998) 309}
  [\href{https://arxiv.org/abs/hep-ph/9711514}{{\ttfamily hep-ph/9711514}}].

\bibitem{Kusenko:1997si}
A.~Kusenko and M.~E. Shaposhnikov, \emph{{Supersymmetric Q balls as dark
  matter}}, \href{https://doi.org/10.1016/S0370-2693(97)01375-0}{\emph{Phys.
  Lett.} {\bfseries B418} (1998) 46}
  [\href{https://arxiv.org/abs/hep-ph/9709492}{{\ttfamily hep-ph/9709492}}].

\bibitem{Kusenko:1997vp}
A.~Kusenko, V.~Kuzmin, M.~E. Shaposhnikov and P.~G. Tinyakov,
  \emph{{Experimental signatures of supersymmetric dark matter Q balls}},
  \href{https://doi.org/10.1103/PhysRevLett.80.3185}{\emph{Phys. Rev. Lett.}
  {\bfseries 80} (1998) 3185}
  [\href{https://arxiv.org/abs/hep-ph/9712212}{{\ttfamily hep-ph/9712212}}].

\bibitem{Kusenko:2001vu}
A.~Kusenko and P.~J. Steinhardt, \emph{{Q ball candidates for selfinteracting
  dark matter}},
  \href{https://doi.org/10.1103/PhysRevLett.87.141301}{\emph{Phys.\ Rev.\
  Lett.} {\bfseries 87} (2001) 141301}
  [\href{https://arxiv.org/abs/astro-ph/0106008}{{\ttfamily
  astro-ph/0106008}}].

\bibitem{Krylov:2013qe}
E.~Krylov, A.~Levin and V.~Rubakov, \emph{{Cosmological phase transition,
  baryon asymmetry and dark matter Q-balls}},
  \href{https://doi.org/10.1103/PhysRevD.87.083528}{\emph{Phys. Rev.}
  {\bfseries D87} (2013) 083528}
  [\href{https://arxiv.org/abs/1301.0354}{{\ttfamily 1301.0354}}].

\bibitem{Ponton:2019hux}
E.~Pont{\'o}n, Y.~Bai and B.~Jain, \emph{{Electroweak Symmetric Dark Matter
  Balls}}, \href{https://doi.org/10.1007/s13130-019-11194-5}{\emph{JHEP}
  {\bfseries 09} (2019) 011}
  [\href{https://arxiv.org/abs/1906.10739}{{\ttfamily 1906.10739}}].

\bibitem{Bai:2019ogh}
Y.~Bai and J.~Berger, \emph{{Nucleus Capture by Macroscopic Dark Matter}},
  \href{https://doi.org/10.1007/JHEP05(2020)160}{\emph{JHEP} {\bfseries 05}
  (2020) 160} [\href{https://arxiv.org/abs/1912.02813}{{\ttfamily
  1912.02813}}].

\bibitem{Bai:2021mzu}
Y.~Bai, S.~Lu and N.~Orlofsky, \emph{{Q-monopole-ball: a topological and
  nontopological soliton}},
  \href{https://doi.org/10.1007/JHEP01(2022)109}{\emph{JHEP} {\bfseries 01}
  (2022) 109} [\href{https://arxiv.org/abs/2111.10360}{{\ttfamily
  2111.10360}}].

\bibitem{Multamaki:2002hv}
T.~Multamaki and I.~Vilja, \emph{{Simulations of Q ball formation}},
  \href{https://doi.org/10.1016/S0370-2693(02)01730-6}{\emph{Phys. Lett. B}
  {\bfseries 535} (2002) 170}
  [\href{https://arxiv.org/abs/hep-ph/0203195}{{\ttfamily hep-ph/0203195}}].

\bibitem{Tsumagari:2009na}
M.~I. Tsumagari, \emph{{Affleck-Dine dynamics, Q-ball formation and
  thermalisation}},
  \href{https://doi.org/10.1103/PhysRevD.80.085010}{\emph{Phys. Rev. D}
  {\bfseries 80} (2009) 085010}
  [\href{https://arxiv.org/abs/0907.4197}{{\ttfamily 0907.4197}}].

\bibitem{Griest:1989cb}
K.~Griest, E.~W. Kolb and A.~Massarotti, \emph{{Statistical Fluctuations as the
  Origin of Nontopological Solitons}},
  \href{https://doi.org/10.1103/PhysRevD.40.3529}{\emph{Phys. Rev. D}
  {\bfseries 40} (1989) 3529}.

\bibitem{Griest:1989bq}
K.~Griest and E.~W. Kolb, \emph{{Solitosynthesis: Cosmological Evolution of
  Nontopological Solitons}},
  \href{https://doi.org/10.1103/PhysRevD.40.3231}{\emph{Phys. Rev. D}
  {\bfseries 40} (1989) 3231}.

\bibitem{Hiramatsu:2010dx}
T.~Hiramatsu, M.~Kawasaki and F.~Takahashi, \emph{{Numerical study of Q-ball
  formation in gravity mediation}},
  \href{https://doi.org/10.1088/1475-7516/2010/06/008}{\emph{JCAP} {\bfseries
  06} (2010) 008} [\href{https://arxiv.org/abs/1003.1779}{{\ttfamily
  1003.1779}}].

\bibitem{Bai:2022kxq}
Y.~Bai, S.~Lu and N.~Orlofsky, \emph{{Origin of nontopological soliton dark
  matter: solitosynthesis or phase transition}},
  \href{https://doi.org/10.1007/JHEP10(2022)181}{\emph{JHEP} {\bfseries 10}
  (2022) 181} [\href{https://arxiv.org/abs/2208.12290}{{\ttfamily
  2208.12290}}].

\bibitem{Volkov:2002aj}
M.~S. Volkov and E.~Wohnert, \emph{{Spinning Q balls}},
  \href{https://doi.org/10.1103/PhysRevD.66.085003}{\emph{Phys. Rev. D}
  {\bfseries 66} (2002) 085003}
  [\href{https://arxiv.org/abs/hep-th/0205157}{{\ttfamily hep-th/0205157}}].

\bibitem{Campanelli:2009su}
L.~Campanelli and M.~Ruggieri, \emph{{Spinning Supersymmetric Q-balls}},
  \href{https://doi.org/10.1103/PhysRevD.80.036006}{\emph{Phys. Rev. D}
  {\bfseries 80} (2009) 036006}
  [\href{https://arxiv.org/abs/0904.4802}{{\ttfamily 0904.4802}}].

\bibitem{Arodz:2009ye}
H.~Arodz, J.~Karkowski and Z.~Swierczynski, \emph{{Spinning Q-balls in the
  complex signum-Gordon model}},
  \href{https://doi.org/10.1103/PhysRevD.80.067702}{\emph{Phys. Rev. D}
  {\bfseries 80} (2009) 067702}
  [\href{https://arxiv.org/abs/0907.2801}{{\ttfamily 0907.2801}}].

\bibitem{Shnir:2011gr}
Y.~Shnir, \emph{{Q-vortices, Q-walls and coupled Q-balls}},
  \href{https://doi.org/10.1088/1751-8113/44/42/425202}{\emph{J. Phys. A}
  {\bfseries 44} (2011) 425202}
  [\href{https://arxiv.org/abs/1101.5366}{{\ttfamily 1101.5366}}].

\bibitem{Nugaev:2014iva}
E.~Nugaev and A.~Shkerin, \emph{{Investigation of Q-tubes stability using the
  piecewise parabolic potential}},
  \href{https://doi.org/10.1103/PhysRevD.90.016002}{\emph{Phys. Rev. D}
  {\bfseries 90} (2014) 016002}
  [\href{https://arxiv.org/abs/1404.3207}{{\ttfamily 1404.3207}}].

\bibitem{Loiko:2018mhb}
V.~Loiko, I.~Perapechka and Y.~Shnir, \emph{{Q-balls without a potential}},
  \href{https://doi.org/10.1103/PhysRevD.98.045018}{\emph{Phys. Rev. D}
  {\bfseries 98} (2018) 045018}
  [\href{https://arxiv.org/abs/1805.11929}{{\ttfamily 1805.11929}}].

\bibitem{Silveira:1995dh}
V.~Silveira and C.~M.~G. de~Sousa, \emph{{Boson star rotation: A Newtonian
  approximation}}, \href{https://doi.org/10.1103/PhysRevD.52.5724}{\emph{Phys.
  Rev. D} {\bfseries 52} (1995) 5724}
  [\href{https://arxiv.org/abs/astro-ph/9508034}{{\ttfamily
  astro-ph/9508034}}].

\bibitem{Kleihaus:2005me}
B.~Kleihaus, J.~Kunz and M.~List, \emph{{Rotating boson stars and Q-balls}},
  \href{https://doi.org/10.1103/PhysRevD.72.064002}{\emph{Phys. Rev. D}
  {\bfseries 72} (2005) 064002}
  [\href{https://arxiv.org/abs/gr-qc/0505143}{{\ttfamily gr-qc/0505143}}].

\bibitem{Kleihaus:2007vk}
B.~Kleihaus, J.~Kunz, M.~List and I.~Schaffer, \emph{{Rotating Boson Stars and
  Q-Balls. II. Negative Parity and Ergoregions}},
  \href{https://doi.org/10.1103/PhysRevD.77.064025}{\emph{Phys. Rev. D}
  {\bfseries 77} (2008) 064025}
  [\href{https://arxiv.org/abs/0712.3742}{{\ttfamily 0712.3742}}].

\bibitem{Kleihaus:2011sx}
B.~Kleihaus, J.~Kunz and S.~Schneider, \emph{{Stable Phases of Boson Stars}},
  \href{https://doi.org/10.1103/PhysRevD.85.024045}{\emph{Phys. Rev. D}
  {\bfseries 85} (2012) 024045}
  [\href{https://arxiv.org/abs/1109.5858}{{\ttfamily 1109.5858}}].

\bibitem{Liebling:2012fv}
S.~L. Liebling and C.~Palenzuela, \emph{{Dynamical Boson Stars}},
  \href{https://doi.org/10.12942/lrr-2012-6}{\emph{Living Rev. Rel.} {\bfseries
  15} (2012) 6} [\href{https://arxiv.org/abs/1202.5809}{{\ttfamily
  1202.5809}}].

\bibitem{Davidson:2016uok}
S.~Davidson and T.~Schwetz, \emph{{Rotating Drops of Axion Dark Matter}},
  \href{https://doi.org/10.1103/PhysRevD.93.123509}{\emph{Phys. Rev. D}
  {\bfseries 93} (2016) 123509}
  [\href{https://arxiv.org/abs/1603.04249}{{\ttfamily 1603.04249}}].

\bibitem{Herdeiro:2019mbz}
C.~Herdeiro, I.~Perapechka, E.~Radu and Y.~Shnir, \emph{{Asymptotically flat
  spinning scalar, Dirac and Proca stars}},
  \href{https://doi.org/10.1016/j.physletb.2019.134845}{\emph{Phys. Lett. B}
  {\bfseries 797} (2019) 134845}
  [\href{https://arxiv.org/abs/1906.05386}{{\ttfamily 1906.05386}}].

\bibitem{Collodel:2019ohy}
L.~G. Collodel, B.~Kleihaus and J.~Kunz, \emph{{Structure of rotating charged
  boson stars}}, \href{https://doi.org/10.1103/PhysRevD.99.104076}{\emph{Phys.
  Rev. D} {\bfseries 99} (2019) 104076}
  [\href{https://arxiv.org/abs/1901.11522}{{\ttfamily 1901.11522}}].

\bibitem{Delgado:2020udb}
J.~F.~M. Delgado, C.~A.~R. Herdeiro and E.~Radu, \emph{{Rotating Axion Boson
  Stars}}, \href{https://doi.org/10.1088/1475-7516/2020/06/037}{\emph{JCAP}
  {\bfseries 06} (2020) 037}
  [\href{https://arxiv.org/abs/2005.05982}{{\ttfamily 2005.05982}}].

\bibitem{Kling:2020xjj}
F.~Kling, A.~Rajaraman and F.~L. Rivera, \emph{{New solutions for rotating
  boson stars}}, \href{https://doi.org/10.1103/PhysRevD.103.075020}{\emph{Phys.
  Rev. D} {\bfseries 103} (2021) 075020}
  [\href{https://arxiv.org/abs/2010.09880}{{\ttfamily 2010.09880}}].

\bibitem{Dmitriev:2021utv}
A.~S. Dmitriev, D.~G. Levkov, A.~G. Panin, E.~K. Pushnaya and I.~I. Tkachev,
  \emph{{Instability of rotating Bose stars}},
  \href{https://doi.org/10.1103/PhysRevD.104.023504}{\emph{Phys. Rev. D}
  {\bfseries 104} (2021) 023504}
  [\href{https://arxiv.org/abs/2104.00962}{{\ttfamily 2104.00962}}].

\bibitem{Gervalle:2022fze}
R.~Gervalle, \emph{{Chains of rotating boson stars}},
  \href{https://doi.org/10.1103/PhysRevD.105.124052}{\emph{Phys. Rev. D}
  {\bfseries 105} (2022) 124052}
  [\href{https://arxiv.org/abs/2206.03982}{{\ttfamily 2206.03982}}].

\bibitem{Siemonsen:2023hko}
N.~Siemonsen and W.~E. East, \emph{{Binary boson stars: Merger dynamics and
  formation of rotating remnant stars}},
  \href{https://doi.org/10.1103/PhysRevD.107.124018}{\emph{Phys. Rev. D}
  {\bfseries 107} (2023) 124018}
  [\href{https://arxiv.org/abs/2302.06627}{{\ttfamily 2302.06627}}].

\bibitem{Kobayashi:1994qi}
Y.~Kobayashi, M.~Kasai and T.~Futamase, \emph{{Does a boson star rotate?}},
  \href{https://doi.org/10.1103/PhysRevD.50.7721}{\emph{Phys. Rev. D}
  {\bfseries 50} (1994) 7721}.

\bibitem{Schunck1996}
F.~E. Schunck and E.~W. Mielke, \emph{Rotating Boson Stars}, pp.~138--151.
\newblock Springer Berlin Heidelberg, Berlin, Heidelberg, 1996.

\bibitem{Schunck:1996he}
F.~E. Schunck and E.~W. Mielke, \emph{{Rotating boson star as an effective mass
  torus in general relativity}},
  \href{https://doi.org/10.1016/S0375-9601(98)00778-6}{\emph{Phys. Lett. A}
  {\bfseries 249} (1998) 389}.

\bibitem{Radu:2008pp}
E.~Radu and M.~S. Volkov, \emph{{Existence of stationary, non-radiating ring
  solitons in field theory: knots and vortons}},
  \href{https://doi.org/10.1016/j.physrep.2008.07.002}{\emph{Phys. Rept.}
  {\bfseries 468} (2008) 101}
  [\href{https://arxiv.org/abs/0804.1357}{{\ttfamily 0804.1357}}].

\bibitem{Radu:2008ta}
E.~Radu and M.~S. Volkov, \emph{{Spinning Electroweak Sphalerons}},
  \href{https://doi.org/10.1103/PhysRevD.79.065021}{\emph{Phys. Rev. D}
  {\bfseries 79} (2009) 065021}
  [\href{https://arxiv.org/abs/0810.0908}{{\ttfamily 0810.0908}}].

\bibitem{Heeck:2020bau}
J.~Heeck, A.~Rajaraman, R.~Riley and C.~B. Verhaaren, \emph{{Understanding
  Q-Balls Beyond the Thin-Wall Limit}},
  \href{https://doi.org/10.1103/PhysRevD.103.045008}{\emph{Phys. Rev. D}
  {\bfseries 103} (2021) 045008}
  [\href{https://arxiv.org/abs/2009.08462}{{\ttfamily 2009.08462}}].

\bibitem{Heeck:2021zvk}
J.~Heeck, A.~Rajaraman, R.~Riley and C.~B. Verhaaren, \emph{{Mapping Gauged
  Q-Balls}}, \href{https://doi.org/10.1103/PhysRevD.103.116004}{\emph{Phys.
  Rev. D} {\bfseries 103} (2021) 116004}
  [\href{https://arxiv.org/abs/2103.06905}{{\ttfamily 2103.06905}}].

\bibitem{Heeck:2021gam}
J.~Heeck, A.~Rajaraman and C.~B. Verhaaren, \emph{{Ubiquity of gauged
  Q-shells}}, \href{https://doi.org/10.1103/PhysRevD.104.016030}{\emph{Phys.
  Rev. D} {\bfseries 104} (2021) 016030}
  [\href{https://arxiv.org/abs/2105.02893}{{\ttfamily 2105.02893}}].

\bibitem{Heeck:2021bce}
J.~Heeck, A.~Rajaraman, R.~Riley and C.~B. Verhaaren, \emph{{Proca Q-balls and
  Q-shells}}, \href{https://doi.org/10.1007/JHEP10(2021)103}{\emph{JHEP}
  {\bfseries 10} (2021) 103}
  [\href{https://arxiv.org/abs/2107.10280}{{\ttfamily 2107.10280}}].

\bibitem{Almumin:2021gax}
Y.~Almumin, J.~Heeck, A.~Rajaraman and C.~B. Verhaaren, \emph{{Excited
  Q-balls}}, \href{https://doi.org/10.1140/epjc/s10052-022-10772-5}{\emph{Eur.
  Phys. J. C} {\bfseries 82} (2022) 801}
  [\href{https://arxiv.org/abs/2112.00657}{{\ttfamily 2112.00657}}].

\bibitem{Nugaev:2019vru}
E.~Nugaev and A.~Shkerin, \emph{{Review of non-topological solitons in theories
  with $U(1)$-symmetry}},
  \href{https://doi.org/10.1134/S1063776120020077}{\emph{J. Exp. Theor. Phys.}
  {\bfseries 130} (2020) 301}
  [\href{https://arxiv.org/abs/1905.05146}{{\ttfamily 1905.05146}}].

\bibitem{PaccettiCorreia:2001wtt}
F.~Paccetti~Correia and M.~G. Schmidt, \emph{{Q balls: Some analytical
  results}}, \href{https://doi.org/10.1007/s100520100710}{\emph{Eur. Phys. J.
  C} {\bfseries 21} (2001) 181}
  [\href{https://arxiv.org/abs/hep-th/0103189}{{\ttfamily hep-th/0103189}}].

\bibitem{Kling:2017mif}
F.~Kling and A.~Rajaraman, \emph{{Towards an Analytic Construction of the
  Wavefunction of Boson Stars}},
  \href{https://doi.org/10.1103/PhysRevD.96.044039}{\emph{Phys. Rev. D}
  {\bfseries 96} (2017) 044039}
  [\href{https://arxiv.org/abs/1706.04272}{{\ttfamily 1706.04272}}].

\bibitem{Kling:2017hjm}
F.~Kling and A.~Rajaraman, \emph{{Profiles of boson stars with
  self-interactions}},
  \href{https://doi.org/10.1103/PhysRevD.97.063012}{\emph{Phys. Rev. D}
  {\bfseries 97} (2018) 063012}
  [\href{https://arxiv.org/abs/1712.06539}{{\ttfamily 1712.06539}}].

\end{thebibliography}\endgroup

\end{document}